\documentclass[twoside, a4paper, notoc, 10pt]{JHEP3}
\usepackage{amsfonts}
\usepackage{amssymb}
\usepackage{comment}
\usepackage{epsfig}
\usepackage{graphicx}
\usepackage{amsmath}
\usepackage{amsxtra}

\newcommand{\bea}{\begin{eqnarray}}
\newcommand{\eea}{\end{eqnarray}}

\newcommand{\be}{\begin{equation}}
\newcommand{\ee}{\end{equation}}

\def\ba{\begin{eqnarray}}
\def\ea{\end{eqnarray}}

\makeatletter
\def\x@arrow{\DOTSB\Relbar}
\def\xlongequalsignfill@{\arrowfill@\x@arrow\Relbar\x@arrow}
\newcommand{\xlongequal}[2]{%
    \ext@arrow 0099\xlongequalsignfill@{#1}{#2}}
\makeatother

\newcommand{\mathbold}[1]{\mbox{\boldmath $#1$}}
\newcommand{\roughly}[1]{\mathrel{\raise.3ex\hbox{$#1$\kern-0.85em
\lower1ex\hbox{$\sim$}}}}

\def\endignore{}
\def\ignore #1\endignore{}

\def\beq{\begin{equation}}
\def\eeq{\end{equation}}
\def\beqa{\begin{eqnarray}}
\def\eeqa{\end{eqnarray}}

\newcommand{\bmat}{\left(\begin{array}}
\newcommand{\emat}{\end{array}\right)}

\def\endignore{}
\def\ignore #1\endignore{}

\def\-{\hphantom{-}}

\def\s2{\frac{1}{2}}

\def\IF{\relax{\rm I\kern-.18em F}}
\def\II{\relax{\rm I\kern-.18em I}}
\def\IP{\relax{\rm I\kern-.18em P}}
\def\IC{\relax{\rm I\kern-.48em C}}
\def\IR{\relax{\rm I\kern-.18em R}}
\def\IK{\relax{\rm I\kern-.20em K}}
\def\IM{\relax{\rm I\kern-.25em M}}

\def\Dsl{\,\raise.15ex\hbox{/}\mkern-13.5mu D}

\def \one{\relax{\rm 1\kern-.26em I}}

\def\d{\mathrm{d}}

\def\({\left(}
\def\){\right)}

\def\a{\alpha}
\def\d{\delta}
\def\aprime{\a^\prime}

\title{A Comment on Continuous Spin Representations of the Poincar\'e Group and Perturbative String Theory }

\author{Anamar\'{\i}a Font,${}^{1}$  Fernando Quevedo${}^{2,3}$  and Stefan Theisen${}^{4}$ \\

\noindent
$^1$ Departamento de F\'{\i}sica, Centro de F\'{\i}sica Te\'orica y Computacional \\
$\phantom{^1}$ Facultad de Ciencias, Universidad Central de Venezuela \\
$\phantom{^1}$ A.P. 20513, Caracas1020-A, Venezuela 

$^2$ Abdus Salam ICTP, Strada Costiera 11, Trieste 34014, Italy

$^3$ DAMTP/CMS, University of Cambridge, Wilberforce Road,
Cambridge CB3 0WA, UK.

$^4$ Max-Planck-Institut f\"ur Gravitationsphysik, Albert-Einstein-Institut, 14476 Golm, Germany 
}

\abstract{We make a simple observation that the massless continuous spin representations of the 
Poincar\'e group are not present in perturbative string theory constructions. This represents
one of the very few model-independent low-energy consequences of these models.\\

{\email{ afont@fisica.ciens.ucv.ve}; \email{F.Quevedo@damtp.cam.ac.uk}; \email{stefan.theisen@aei.mpg.de}}}

\preprint{DAMTP-2013-10}

\begin{document}

\tableofcontents

\bigskip

\section{Introduction and Conclusions}

String theory, just like standard relativistic quantum field theories, has  very few model 
independent consequences at low energies. In quantum field theory we can name the existence of 
anti-particles, the CPT theorem, the running of couplings in terms of the renormalisation group 
and the identity of all particles of the same type. 
String theory, for vacua with non-compact dimensions,   
`predicts' gravity and at least one neutral scalar, the dilaton, antisymmetric tensors of different ranks 
and usually also charged matter, and supersymmetry (see for instance \cite{witten}). 

By the same nature of a theory (string theory or otherwise) with a naturally large energy scale to 
address the issue of quantum gravity,  it is very difficult to identify model independent low-
energy implications subject to experimental  verification which can put to test the theory and not 
just particular models or scenarios.

The purpose of this note is to make a simple but general remark. We point out a low-energy 
consequence of all string constructions, that is the absence of massless continuous spin 
representations (CSR) of the Poincar\'e group \cite{Wigner}. This fact  has no straightforward explanation 
within standard particle physics field theoretical analysis and is consistent with all experiments 
so far since particles fitting into those representations have not been found in nature; 
see however \cite{toro} or a recent discussion of their phenomenology.

One of the most elegant theoretical developments in particle physics, pioneered by Wigner, 
is the description of one-particle states in terms of unitary representations of the four-
dimensional Poincar\'e group \cite{Wigner} (see also \cite{weinberg}).

One-particle states  are  classified according to the quantum numbers of the invariant Casimir 
operators $C_1=P^\mu P_\mu$ and $C_2=W^\mu W_\mu$ with $P^\mu$ and  
$W^\mu=\epsilon^{\mu\nu\rho\sigma}P_\nu M_{\rho\sigma}$ the momentum and Pauli-Ljubansky vectors 
respectively and $M_{\rho\sigma}$ the Lorentz generators.  $C_1$ and $C_2$ label the 
representation in terms  of their eigenvalues that  essentially correspond to mass $m$  and spin 
$J$ in a representation with fixed momentum $p_\mu$.  

The representations differ according to whether $C_1$ is positive, zero or negative.
For massive particles ($C_1>0$) the remaining space-time quantum numbers come from the fact that 
the stabilising or Little  Group in four-dimensions is $SO(3)$, the subgroup of the Poincar\'e 
group leaving invariant a state in its rest mass frame described by a four-momentum $p=(m,0,0,0)$. 
The corresponding states are the different spin states of the multiplet. This defines a particle 
in terms of quantum numbers $|m,J; p_\mu, s\rangle$ with $s=-J, -J+1, \cdots, J$ and
$p^2=m^2$. 

For massless particles ($C_1=0$) \footnote{The case $C_1<0$ corresponds to tachyonic states 
that are usually a signature of instability. In supersymmetric string theories this particle 
is projected out of the spectrum, although being the ground state of the quantisation 
it has played important roles in the understanding of branes and with potential cosmological 
implications \cite{sen}.} the momentum can be written $p=(E,0,0,E)$ 
and the corresponding Little Group is not only the naively guessed $SO(2)$ but actually 
the whole Euclidean group in two dimensions $E_2$ or $ISO(2)$.
This complicates matters since this group has infinite dimensional unitary representations, 
known as continuous spin representations (CSR), that would correspond to a continuous 
spin-like label on the elementary particles, something that has not been observed in nature.

A standard way to proceed is to simply restrict to the finite dimensional representations that 
correspond to those of $SO(2)$. This defines helicity $\lambda$, as the good quantum number 
which is quantised in half integers.  Since the two Casimirs vanish in the reference frame  
defined by $p=(E,0,0,E)$ all observed massless particles 
\footnote{Recall that in the standard model all known particles, 
except the  Higgs particle itself, are described by 
massless states and the massive ones acquire their mass via the Higgs effect.} 
are then labelled only by $p_\mu$ and $\lambda$:
 $|p_\mu, \lambda\rangle$ with $\lambda= 0, \pm 1/2, \pm 1, \cdots$. But there is no 
satisfactory explanation why to restrict only to representations of $SO(2)$ instead of the full 
$E_2$. Contrary to the massive case for which matter fields fit into generic representations of 
$SO(3)$, there are massless representations (infinite dimensional) that are allowed by the basic 
principles of special relativity and quantum mechanics but do not seem to be realised in nature. 
A theoretical understanding of this fact is needed.

Over the years the continuous spin representations have been discussed in several different 
contexts (see for instance \cite{brink} and references therein) and attempts have been made to
describe them in terms of quantum field theoretical interactions, but without much success. The
question of their relevance becomes even stronger in the description of higher 
dimensional theories, such as ten and eleven dimensional supergravities, for which the 
argument that they have not been observed in nature does not  directly apply. 
Therefore we may wonder if either these representations exist and may have an 
important role to play in a fundamental theory or
if the structure of the fundamental theory  may provide a first-principles explanation 
of why these particles are not observed in nature.

In this note we would like to address the relevance of string theory for the existence or 
not of continuous spin representations. One may ask if these states could be present in string theory. 
In perturbative string theory we can observe  an obstruction since in the standard quantisation, particles of different 
masses and spin are in the same multiplet 
in the sense that upon application of the creation and annihilation operators one 
relates particles of different masses. It is then clear that if the massive representations 
do not carry a continuous label the massless states should not carry it either. 
Then the continuous spin representations are not present in perturbative string constructions. 
This argument can be turned into a model-independent prediction of string constructions 
at the same level as the other two general `predictions'  of the theory, the presence 
of gravity and supersymmetry.  It can be said that if these states are detected
experimentally, all string theory constructions known so far would be ruled out. 
From the perspective of the CSR's string theory  provides a straightforward explanation of 
why the relevant part of the Little group for massless states is $O(D-2)$ for a $D$ 
dimensional theory instead of the full $ISO(D-2)$.

The fact that these representations are not realised in perturbative string theory does not preclude 
a potential explanation in terms of field theory itself (see for 
instance \cite{schroer}). In principle it may be the that interacting 
field theories for these states may prove inconsistent \footnote{However a series of 
recent articles indicate the opposite, making these states much more interesting and physical 
than usually appreciated \cite{toro}.}. But in perturbative string theory the statement is much 
cleaner since these representations simply do not appear in the standard way of quantising 
the theory. It may also be argued that in terms of the standard gravity/gauge theory
correspondence, if these states are not present on the gravity side they should not be 
present in any field theory with a string dual. Furthermore if there exist  
interacting field theories for these states, they should belong to the swampland of  
the string landscape \cite{vafa}.

Finally it has been suggested that CSR's could be realised in the zero tension limit of 
string theory in which the infinite tower of massive states collapses to zero mass (see for instance \cite{savvidy}). 
This could be very interesting, but even if true  it does not correspond to the standard 
string constructions that upon compactifications lead to realistic low energy effective 
field theories. 
In the remainder of this note we will make the argument for the absence of CSR explicit  
by actually computing the action of the extra generators on the string states and 
show that they vanish.

\section{Review of CSRs  in $\mathbold{d}$ dimensions}
\label{Review}

The generators of the Poincar\'e group satisfy the well known algebra
\beqa
\left[M^{\mu\nu}, M^{\a\beta}\right]& = & 
i\left( \eta^{\mu\a} M^{\nu\beta} - \eta^{\nu\a} M^{\mu\beta} 
+ \eta^{\nu\beta} M^{\mu\a} - \eta^{\mu\beta} M^{\nu\a}\right) \\
\left[M^{\mu\nu}, P^\a\right]& = & i\left( \eta^{\mu\a} P^\nu 
- \eta^{\nu\a} P^\mu \right) \quad; \quad  \left[P^\a, P^\beta\right] = 0
\label{palg}
\eeqa
where the Greek indices run from $0$ to $d-1$. In light cone frame the 
momenta are the $P^i$ together with
\beq
P^{\pm} = \frac1{\sqrt2}\left(P^0 \pm P^{d-1}\right)
\label{ppm}
\eeq
while the Lorentz generators split into the $M^{ij}$ and
\beq
M^{+-}=M^{d-1\, 0} \quad ; \quad M^{\pm i}= \frac1{\sqrt2}\left(M^{0i} \pm M^{d-1\, i}\right)
\eeq
where $i=1,\cdots,d-1$.

For massless particles the representative momentum is $p^\mu=(E, 0,\cdots,E)$, 
so that $p^+=\sqrt2 E$ while $p^-=0$ and $p^i=0$. Therefore, the generators 
of the little group are $M^{ij}$ and $\Pi^i\equiv M^{-i}$
which leave the representative momentum invariant \cite{bekaert}. 
These generators satisfy the $ISO(d-2)$ algebra
\beqa
\left[M^{ij}, M^{kl}\right]& = & i\left( \d^{ik} M^{jl} - \d^{jk} M^{il} 
+ \d^{jl} M^{ik} - \d^{il} M^{jk}\right) \\
\left[M^{ij}, \Pi^k\right]& = & i\left( \d^{ik} \Pi^j - \d^{jk} \Pi^i \right) \quad; \quad  
\left[\Pi^i, \Pi^j\right] = 0
\label{isoalg}
\eeqa
The helicity representation is obtained when $\Pi^i\equiv 0$ on the states and the algebra 
reduces to that of $SO(d-2)$.
The CSR representations are obtained for $\sum_i(\Pi^i)^2 \not= 0$. 

To see what this condition implies, consider for simplicity $d=4$, i.e. $i=1,2$. With the defintions
 $M^{12}=J_3$ and $\Pi^{\pm}=\Pi^1\pm i \Pi^2$, the algebra 
\eqref{isoalg} becomes $[\Pi^+,\Pi^-]=0$ and 
$[J_3,\Pi^\pm]=\pm\Pi^\pm$. Given an eigenstate of $J_3$ with integer or half-integer
eigenvalue $\sigma$, i.e. $J_3|\sigma\rangle=\sigma|\sigma\rangle$, we can create states 
$|\sigma\pm n\rangle=(\Pi^\pm)^n|\sigma\rangle$ with eigenvalues $(\sigma\pm n$) for any 
positive integer $n$. Going to Fourier space we can construct states 
$|\theta\rangle=\sum_\sigma e^{-i\sigma\theta}|\sigma\rangle$ which simultaneously diagonalize 
$\Pi^\pm$, i.e. $\Pi^{\pm}|\theta\rangle=e^{\pm i\theta}|\theta\rangle$, whereas a rotation acts as
$e^{-i\alpha J_3} |\theta\rangle =  |\theta+ \alpha \rangle$.   
It is this continuous label $\theta\in[0,2\pi)$ which  gives rise to the name 
`continuous spin representation'.

To prepare for the discussion of the string spectrum, we first 
review the point particle, mainly following ref. \cite{brink,ramond} which are, in 
parts, based on \cite{bacry}. 

In systems with reparametrization invariance, such as the point particle, 
it is often convenient to work in the light-cone 
gauge in which $x^+$ is fixed. 
In this case $p^-$ is no longer a conjugate momentum. 
Instead $p^-$ is determined to be
\beq
p^- = \frac{p^i p^i + m^2}{2 p^+}
\label{pmin}
\eeq
which follows from the on-shell constraint for a particle of mass $m$. 
The conjugate pairs satisfy $[x^i,p^j]=i \d^{ij}$, and $[x^-,p^+]=-i$. 
The translation operators are 
$P^+=p^+$ and $P^i=p^i$, while
$P^-=p^-$ is the light-cone Hamiltonian.

The Lorentz generators can be decomposed into orbital and spin parts. Concretely,
\beq
M^{ij} = x^i p^j - x^jp^i + S^{ij}
\label{mijspin} 
\eeq
where the $S^{ij}$ are $SO(d-2)$ generators. Moreover,
\beq
M^{+i} = -x^i p^+ \quad ; \quad M^{+-} = -\frac12 \left\{x^-,p^+\right\}
\label{mplus}
\eeq
In $d=4$ it can be shown explicitly that $S^{+i}=0$ and $S^{+-}=0$ \cite{ramond}. 
In general one can show that the above $M^{+-}$ and $M^{+j}$, together with  
\beq
M^{-i} =x^-p^i -\frac12 \left\{x^i,p^-\right\} + \frac1{p^+}\left(T^i - p^j S^{ij}\right)
\label{mminus}
\eeq
satisfy the Lorentz algebra. 
The $T^i$ are $SO(d-2)$ vectors, i.e. 
$\left[S^{ij},T^k\right]=i\left(\d^{ik} T^j - \d^{jk} T^i \right)$, and
further satisfy
\beq
\left[T^i,T^j\right]=i m^2 S^{ij}
\label{ttcom}
\eeq
In $d=4$ the $T^i$ are constructed explicitly and shown to verify these properties \cite{ramond}. In general they
follow imposing that the $M^{-i}$ in (\ref{mminus}) fulfill the Lorentz algebra. In particular, (\ref{ttcom})
guarantees that $\left[M^{-i},M^{-j}\right]=0$.  

When $m=0$ the $T^i$ commute among themselves and become the translation operators of $ISO(d-2)$. 
Thus, in the helicity representation $T^i =0$ whereas in the CSR $T^i \not = 0$. 

In \cite{brink} the authors generalize the above discussion to continuous spin 
representations of the supersymmetry algebra. 
The question whether these CSRs can be incorporated into an 11-dimensional supergravity
theory is possibly of interest in relation to M-theory and the continuous excitation spectrum 
of the membrane \cite{deWit}.\footnote{We thank M. Green for mentioning this possibility.}

\section{CSRs in String Theory}

In the (open) bosonic string, in the critical dimension $d=26$, the translation operators 
$\Pi^i\equiv M^{-i}$ in the
light cone gauge are given by \cite{grt, ggrt} \footnote{We mostly use the notation 
of \cite{zwiebach}.}
\beq
\Pi^i=x_0^- p^i - \frac12\left\{x_0^i,p^- \right\} - i\sum_{n=1}^\infty \frac1{n}\left(\a_{-n}^-\a_n^i - \a_{-n}^i\a_n^-\right)
\label{mmi}
\eeq
where $[x_0^i,p^j]=i\d^{ij}$, $[x_0^-,p^+]=-i$, $[\a_m^i,\a_{-n}^j]=m\d^{ij}\d_{m,n}$, and
\beqa
p^- &=& \frac1{2\aprime p^+}\left(\aprime p^i p^i+\sum_{n=1}^\infty n\a_{-n}^i\a_n^i-1\right) \\ 
\a_n^- & = & \frac1{\sqrt{2\aprime} p^+} \sum_{p=-\infty}^\infty \a_{n-p}^i\a_p^i \quad ; \quad 
\a_{-n}^- = \left(\a_n^-\right)^\dagger
\label{opminus}
\eeqa
Normal ordering is understood. Notice that $[x_0^i,p^-]=ip^i/p^+$. 

The $\a_n^i$, $n\ge 1$, annihilate the vacuum $|p^+, \vec{p}\, \rangle$. 
The massless states are ($\vec p=\{p^i\}$)
\beq
|j\rangle \equiv   \a_{-1}^j |p^+, \vec{p}\, \rangle
\label{zeromass}
\eeq
which transform as a vector of $SO(d-2)$. This indicates that the 
little group is $SO(d-2)$. By consistency
we then expect the $\Pi^i$ to be zero acting on these states. 
Using $\a_0^i = \sqrt{2\aprime} p^i$ we find
\beq
p^-|j\rangle = \frac{\vec{p}^2}{2p^+} |j\rangle \quad ; \quad
\a_{-1}^i \a_1^-|j\rangle = \frac{p^j}{p^+} |i\rangle \quad ; \quad
\a_{-1}^- \a_1^i|j\rangle = \d^{ij} \frac{p^k}{p^+} |k\rangle
\label{opmact}
\eeq 
Thus, it follows that $\Pi^i|j\rangle = 0$ when the transverse momentum of the 
massless states verifies $p^i=0$.  
 
In light-cone gauge it is actually more convenient to employ the decomposition 
of the Lorentz generators into orbital and spin parts. 
In the bosonic string the $M^{ij}$ are in fact written as in (\ref{mijspin}) with
the spin piece given by
\beq
S^{ij}= - i\sum_{n=1}^\infty \frac1{n}\left(\a_{-n}^i\a_n^j - \a_{-n}^j\a_n^i\right)
\label{mijstring}
\eeq
Furthermore, comparing (\ref{mminus}) and (\ref{mmi}), 
and using the above expression for $S^{ij}$, we obtain
\beq
T^i=i\sum_{n=1}^\infty \frac1{n}\left[ \a_{-n}^i\left( p^+\a_n^- - p^j\a_n^j\right) - 
\left(p^+\a_{-n}^- -\a_{-n}^j\right)\a_n^i  \right]
\label{opts}
\eeq
It can be shown that these $T^i$'s satisfy (\ref{ttcom}) with $m^2$ replaced 
by the mass operator of the open bosonic string .

Since the massless states $|j\rangle$ belong in the helicity representation we again 
expect that the $T^i$ are zero acting on these states. Indeed, using the last 
two results in (\ref{opmact}), we find $T^i|j\rangle = 0$ and this holds for all $p^i$. 
This is a consistency check that the CSRs do not appear in the perturbative string spectrum.

Here we have only considered the 
contribution to the Lorentz generators from the oscillators which arise in the 
quantization of the non-compact space-time dimensions. In the critical dimension
this is all there is. In the compactified theory there are 
also contributions from the internal CFT but they will not change the argument and 
result. The same is true for the extension to the fermionic string. 

We have observed that using the standard rules for the spectrum of perturbative string models, 
continuous spin representations do not appear. It would be interesting to study  if   
these representations could be present in the full-fledged string theory. 

\bigskip\medskip

\noindent
{\large{\bf Acknowledgements}}

\smallskip

\noindent
F.Q. acknowledges conversations and email exchange over the years with: Nima Arkani-Hamed, Cliff Burgess, 
Freddy Cachazo, Shanta de Alwis, Michael Green,  Anshuman Maharana, Juan Maldacena, Joe Polchinski, Pierre Ramond, 
Seif Randjbar-Daemi and Barton Zwiebach, as well as the 
students of the Part III  Supersymmetry and Extra Dimensions course in Cambridge.
A.F. and S.T. thank the Newton Institute for hospitality in 2002 when this project was
conceived, and ICTP in 2012 where it was developed. 
F.Q. thanks Ospedale Cattinara for `hospitality' in 2012, and the ICTP-SAIFR members for 
hospitality and interesting discussions during the last stages of this project.

\end{document}